\documentclass[conference]{IEEEtran}

\usepackage[latin9]{inputenc}
\usepackage{graphicx}
\usepackage{subcaption}
\usepackage{subfig}
\usepackage{amsmath,amssymb}
\usepackage{color}
\usepackage[nocompress]{cite}
\ifCLASSINFOpdf
\else
\fi
\DeclareMathOperator*{\argmin}{argmin}

\newcommand{\bx}{\ensuremath{\mathbf{x}}}
\newcommand{\bX}{\ensuremath{\mathbf{X}}}
\newcommand{\by}{\ensuremath{\mathbf{y}}}
\newcommand{\bh}{\ensuremath{\mathbf{h}}}

\begin{document}
%
\title{A Large-Scale Exploration of Factors Affecting Hand Hygiene Compliance Using Linear Predictive Models}

\author{
\IEEEauthorblockN{Michael T. Lash\IEEEauthorrefmark{1} MS, Jason Slater\IEEEauthorrefmark{2} BS,\\
Philip M. Polgreen\IEEEauthorrefmark{3} MD MPH, and Alberto Maria Segre\IEEEauthorrefmark{1} PhD}\\
\IEEEauthorblockA{\IEEEauthorrefmark{1}Department of Computer Science, The University of Iowa, Iowa City, IA 52242,\\ Email: \{michael-lash, alberto-segre\}@uiowa.edu}
\IEEEauthorblockA{\IEEEauthorrefmark{2}Gojo Industries, Inc., Akron, OH 44309-0991, Email: slaterj@gojo.com} \IEEEauthorblockA{\IEEEauthorrefmark{3}Department of Internal Medicine, The University of Iowa, Iowa City, IA 52242, Email: philip-polgreen@uiowa.edu}}


%


\maketitle

\begin{abstract}
This large-scale study, consisting of 24.5 million hand hygiene opportunities
spanning 19 distinct facilities in 10 different states, uses linear
predictive models to expose factors that may affect hand hygiene
compliance. We examine the use of features such as temperature, relative
humidity, influenza severity, day/night shift, federal
holidays and the presence of new residents in predicting daily hand
hygiene compliance. The results suggest that colder temperatures
and federal holidays have an adverse effect on hand hygiene compliance
rates, and that individual cultures and attitudes regarding hand
hygiene exist among facilities.
\end{abstract}



%
\IEEEpeerreviewmaketitle

\section{Introduction}
Healthcare associated infections represent a major cause of morbidity
and mortality in the United States and other countries \cite{Klevens2007}.
Although many can be treated, these infections add greatly to healthcare
costs \cite{Roberts2010}. Furthermore, the emergence of multidrug
resistant bacteria have greatly complicated treatment of healthcare
associated infections \cite{Roberts2009}, making the prevention of
these infections even more important. One of the most effective interventions
for preventing healthcare associated infections is hand hygiene \cite{Boyce2002}.
Yet, despite international programs aimed at increasing hand hygiene \cite{Boyce2002,Allegranzi2012,Pittet2009},
rates remain low, less than 50\% in most cases \cite{Boyce2002,Pittet2009,Hass2007}.

Because of the importance of hand hygiene in preventing healthcare
associated infections, infection control programs are encouraged to
monitor rates to encourage process improvement \cite{Boyce2009,Pittet2009,JCO}.
In most cases, hand hygiene monitoring is done exclusively by human
observers, which are still considered the gold standard for monitoring \cite{Hass2007}.
Yet, human observations are subject to a number of limitations. For example, human observers incur high
costs and there are difficulties in standardizing the elicited observations.
Also, the timing and location of observers can greatly affect the
diversity and the quantity of observations \cite{Fries12,Sharma2012}. Furthermore,
the distance of observers to healthcare workers under observation
and the relative busyness of clinical units can adversely affect the
accuracy of human observers \cite{Sharma2012}. The presence of human
observers may artificially increase hand hygiene rates temporarily
as the presence of other healthcare workers can induce peer effects
to increase rates \cite{Eckmanns2006,Monsalve2014}. Finally, the number of human observations possible is quite small
in comparison to the number of opportunities \cite{Hass2007,Eckmanns2006}. Several automated approaches
to monitoring have been proposed \cite{Boscart2008,Boyce2009,Venkatesh2008,Polgreen2010}.
Many of these measure hand hygiene upon entering and leaving a patient's
room. The subsequent activation of a nearby hand hygiene dispenser is
recorded as a hand hygiene opportunity fulfilled whereas, if no such
activation is observed, the opportunity is not satisfied. Such approaches,
while not capturing all five moments of hand hygiene, do provide an
easy and convenient measure of hand hygiene compliance. With automated
approaches becoming more common, a more
comprehensive picture of hand hygiene adherence should emerge, providing new insights into why healthcare workers
abstain from practicing hand hygiene.

\section{Data and Methods}

\subsection{Hand Hygiene Event Data}

Our hand hygiene event data is a proprietary dataset provided by Gojo
Industries. The data were obtained from a number of installations
consisting of \textit{door counter sensors}, which increment a counter
anytime an individual goes in or out of a room, and \textit{hand hygiene sensors}, which increment a counter when soap or alcohol rub are dispensed.
Additional supporting technology was also installed to collect and
record timestamped sensor-reported counts. In this paper, we will
use the term \textit{dispenser event} to designate triggering and
use of an instrumented hand hygiene dispenser and \textit{door event} to designate the triggering of a counter sensor located on one
of the instrumented doors.

A total of 19 facilities in 10 states were outfitted with sensors;
because of privacy concerns, we provide only the state and CDC Division
for each. The facilities comprise a wide range of geographies, spanning
both coasts, the midwest, and the south. A total of 1851 door sensors
and 639 dispenser sensors reported a total of 24,525,806 door events
and 6,140,067 dispenser events, beginning on October 21, 2013 and
ending on July 7, 2014. Each facility contributed an average of 172.3
\textit{reporting days}, making this study the largest investigation
of hand hygiene compliance to date (i.e., larger than the 13.1 million
opportunities reported in \cite{Dai2014}). Assuming each door event
corresponds to a hand hygiene opportunity, we compute an estimated
intra-facility compliance rate of 25.03\%, in line with if not just
below the reported low-end rate found in \cite{JarrinTejada2015}.

The original data, consisting of timestamped counts reported from
individual sensors over short intervals, were re-factored to support
our analysis. First, data from each sensor were binned by timestamp,
$t$, into 12 hour intervals, corresponding to traditional day and
night shifts, as indicated by an additional variable, $nightShift$, defined
as follows:
\[
nightShift=\begin{cases}
1 & t\textrm{ }\epsilon\textrm{ }[7\text{pm},6\text{:}59\text{am}]\\
0 & t\textrm{ }\epsilon\textrm{ }[7\text{am},6\text{:}59\text{pm}]
\end{cases}
\]
Second, door and dispenser counts were aggregated based on day and
night shift so as to produce a series of records. For each such record we
compute \textit{hand hygiene compliance}, or just \textit{compliance},
by dividing the number of reported dispensed events by the number
of door events: 
\[
compliance=\frac{\#\text{ } dispenser}{\# \text{ } door}
\]
Such a definition of compliance assumes that each door event corresponds
to a single \textit{hand-hygiene opportunity} and each dispenser event corresponds to a single \textit{hand-hygiene event} whereas, in reality, a health
care worker might well be expected to perform hand hygiene more than
once per entry, resulting in rates that exceed one, if only slightly. This estimator also ignores the placement of doors with respect to
dispensers: multiple dispensers may well be associated with a single
doorway, and some dispensers may be in rooms having multiple doors.
Adding new dispensers will raise apparent compliance rates, while
adding new door sensors will appear to reduce compliance. Even so,
when applied consistently and if system layouts are fixed, this estimator
is a reasonable approximation of true hand hygiene compliance, and
supports sound comparisons within a facility (but not across facilities).

Because malfunctioning sensors or dead batteries can produce outliers
(i.e., very low or very high values), records with fewer than 10 door
or dispenser events reported per day (possibly indicating an installation
undergoing maintenance), zero compliance, or compliance values greater
than 1 were removed prior to analysis (at the cost of possibly excluding
some legal records). The remaining data consists of 5308 shifts from
the original 5647 records, having \textcolor{blue}{21,273,980} hand hygiene opportunities and \textcolor{red}{5,296,749} hand hygiene events (see Table \ref{tab:facility_descript}).
\begin{table}[h]
	\centering
	\begin{tabular}{|l|c|c|c|c|c|}
		\hline
		\textbf{Facility} & \textbf{State} & \textbf{CDC Div} & \textbf{Tot Disp} & \textbf{Tot Door} & \textbf{Days Rep} \\
		\hline
		91 & OH & ENC & 234292 & 518772 & 252 \\
		\hline
		101 & OH & ENC & 350901 & 2021665& 260 \\
		\hline
		105 & TX & WSC & 238899 & 1940024 & 260 \\
		\hline
		119 & MN & WNC & 123877 & 242939 & 156 \\
		\hline
		123 & TX & WSC & 325618 & 1112198 & 243 \\
		\hline
		127 & NM & Mnt & 1306855 & 4546171 & 260 \\
		\hline
		135 & OH & ENC & 125731 & 264331 & 258 \\
		\hline
		144 & CA & Pac & 398961 & 1744642 & 260 \\
		\hline
		145 & CA & Pac & 567096 & 2073566 & 260 \\
		\hline
		147 & CA & Pac & 500979 & 2462900 & 260 \\
		\hline
		149 & CA & Pac & 590708 & 2306392 & 260 \\
		\hline
		153 & CT & New E & 169564 & 603482 & 208 \\
		\hline
		155 & NY & M-At & 171275 & 619507 & 117 \\
		\hline
		156 & NC & S-At & 4381 & 38200 & 15 \\
		\hline
		157 & OH & ENC & 39455 & 313396 & 101 \\
		\hline
		163 & OH & ENC & 344 & 10233 & 5 \\
		\hline
		168 & PA & M-At & 30421 & 86909 & 20 \\
		\hline
		170 & IL & ENC & 112604 & 353631 & 47 \\
		\hline
		173 & OH & ENC & 4788 & 15122 & 32 \\
		\hline
		Total & 10 & 8 & \textcolor{red}{5296749} & \textcolor{blue}{21273980} & 3274 \\
		\hline

	\end{tabular}
	
	\caption{Descriptive statistics for all reporting facilities in terms of state, CDC division, hand hygiene events, people events, and reporting days. \label{tab:facility_descript}}
\end{table}
\subsection{Feature Definitions}

In this subsection we define the features (factors) that will be examined, and how each is derived.

\begin{figure}[b]
	\centering
	\includegraphics[scale=1.50]{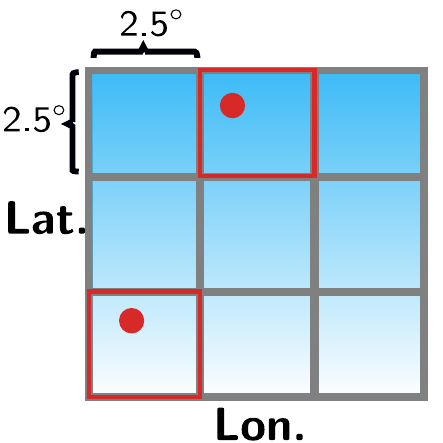}
	\caption{Assigning (\textcolor{red}{red} box) NOAA weather data, reported in terms of a geographic grid, to health care facilities (\textcolor{red}{red} dots), where the \textcolor{cyan}{blue} color gradient might represent temperature. \label{fig:weather_assign}}
\end{figure}

\subsubsection{Local Weather Data}

Because health care workers frequently cite skin dryness and irritation
as a factor in decreased compliance (particularly in cold weather
months where environmental humidity is reduced), we associate daily
air temperature and relative humidity to each timestamped record based
on each facility's reported zip code. Spatially assimilated weather
values ($\sigma=0.995$) for the entire globe were obtained from the
National Oceanic and Atmospheric Administration (NOAA) \cite{Kalnay1996}.
Given in terms of grid elements (a tessilation of bounding boxes covering
$2.5^{\textdegree}$ latitude by $2.5^{\textdegree}$ longitude), the
world is thus defined as a 144 by 73 grid having 10512 distinct grid elements.
Weather data are available at a fine level of temporal granularity
(on the order of 4 times daily for each grid unit) for the entire
period of interest. The geographical assignment of weather data was obtained
by first mapping each facility's numerical zipcode to
the zipcode's centroid (2010 US Census data), given by (latitude,longitude), which was subsequently 
mapped to the matching NOAA grid element. An example of this assignment can be observed in Figure \ref{fig:weather_assign}. We associate weather information from the 6am reporting hour with
records corresponding to traditional day shifts (7am-6:59pm) and use
the 6pm reporting hour for traditional night shifts (7pm-6:59am).

\subsubsection{Influenza Severity}

We conjecture that the local \textit{severity} of
common seasonal diseases, such as influenza, may also affect hand hygiene
compliance rates. We define influenza severity as the number of
influenza-related deaths relative to all deaths over a specified time
interval.

Influenza severity data were obtained from the CDC's \textit{Morbidity and Mortality Weekly Report} (MMWR), which also reports data at weekly
temporal granularity. Rather than reporting data by CDC region, however, data
are provided by \textit{reporting city} (one of 122 participating cities,
mostly large metropolitan areas). We map each facility in our dataset to the closest reporting city in order to associate the appropriate
severity value to each record. In other words

\[
repCity = \argmin\{\text{dist}(\text{facility},\text{city}_i): i=1,\dots,122\}
\]

where $\text{dist}(\text{fac},\text{city}) \triangleq \left \lVert (\text{fac}_{lat},\text{fac}_{lon}), (\text{city}_{lat},\text{city}_{lon}) \right \rVert_2$, the Euclidean distance between two entities
given in terms of (lat, lon) coordinates. Eight of 19 facilities were located
in a reporting city (i.e., dist$=0$). The remaining 11 facilities were mapped to
a reporting city that was, on average, 66.2 miles away (only 3 of 19
facilities were mapped to a reporting city further than this average, with the largest distance being 142 miles).


\subsubsection{Temporal Factors}

We also conjecture that external factors associated with specific
holidays or events may affect hand hygiene compliance rates. Holidays
may change staffing rates or affect healthcare worker behaviors in
various ways. The number of visitors (affecting door
counter rates) may also be greater than during regular weekdays. Holidays
such as the 4th of July are often associated with alcohol-related
accidents, and may increase health care facility workloads. Such
factors may also be observable during weekends.

We define a new variable $holiday$ that reflects whether a given
shift occurs on one of the 10 federal holidays (New Year's Eve, Martin Luther King
Day, President's Day, Memorial Day, the 4th of July, Labor Day, Columbus
Day, Veteran's Day, Thanksgiving or Christmas):

\[
holiday=\begin{cases}
0 & t\notin\left\{ holidays\right\} \\
1 & t\in\left\{ holidays\right\} 
\end{cases}
\]
Similarly, in order to ascertain the impact of weekends on compliance,
we define a new variable $weekday$ as follows:

\[
weekday=\begin{cases}
0 & t\in\left\{ Sat,Sun\right\} \\
1 & t\in\left\{ Mon,Tues,Weds,Thurs,Fri\right\} 
\end{cases}
\]
A related concept is the presence of new resident physicians, who
traditionally start work the first of July. We define a new variable
that corresponds with this time period in order to see if the data
reveal the presence of a July effect:

\[
JulyEffect=\begin{cases}
0 & t\notin July_{1-7}\\
1 & t\in July_{1-7}
\end{cases}
\]

\subsection{Exploring Factors Affecting Hand Hygiene}

\subsubsection{$M5$ Ridge Regression for Feature Examination}

With covariates defined and associated with the collected sensor data,
we 
wish to build a linear hypothesis $\bh$ that \textbf{(a)} accurately estimates hand hygiene and \textbf{(b)} reports the direction and degree of effect of our defined features.

In accomplishing \textbf{(b)} we bear in mind two things:
\begin{enumerate}
	\item[\textbf{(1)}] There may be multi-collinearity among features, which may adversely affect the output.
	\item[\textbf{(2)}] That \textbf{(a)} and \textbf{(b)} may be at odds with one another; i.e.,~obtaining good predictions may entail discarding some prediction-inhibiting features for which we would like to obtain effect estimates (in practice, we find that this is not actually the case).
\end{enumerate}
	

Therefore, we propose an \textit{$M5$ Ridge Regression for Feature Examination} method designed to accomplish \textbf{(a)} and \textbf{(b)}, while bearing \textbf{(1)} and \textbf{(2)} in mind. This method is given by

\begin{align}
\label{eq:ridge}
\begin{array}{lll}
\bh^* = & \argmin \limits_{\bh \in \mathcal{H}_l} & \left \lVert \Lambda(\mathbf{X}) \bh-\by \right \rVert_{2}^{2} + \lambda \left \lVert \bh \right \rVert_{2}^{2} \\
& \text{s.t.} &\rho(h_j) \leq .05 \text{ }\forall \text{ }j
\end{array}
\end{align}


where $\mathbf{X} \in \mathbb{R}^{n \times p}$ is a design matrix, $\bh$ is the hypothesis, $\by$ is the target vector consisting of compliance rates in which a particular $y_i \in [0,1]$, $\lambda$
is a regularization term, $\left \lVert \cdot \right \rVert_2$ is the $\ell_2$-norm, and $\rho(\cdot)$ is a function that reports the p-value of a hypothesis term (this constraint is ensured via sequential backwards elimination \cite{Draper1966}). The function $\Lambda(\mathbf{X})$ can be defined as

\begin{align}
\label{eq:m5}
\Lambda(\mathbf{X}) \triangleq \argmin\{\mathbf{t} \in T_{\mathcal{H}_{l}}\}
\end{align}
where $\mathbf{t}$ is hypothesis selected from a tree of hypotheses constructed using the $M5$ method \cite{Quinlan1992}. Effectively, \eqref{eq:m5} only reduces the $p$ dimension, acting as a feature selection method, and having no bearing on the $n$ dimension.

There are a few benefits of the above method worth pointing out. First, the hypothesis class $\mathcal{H}_l$ is linear and common to both \eqref{eq:ridge} and \eqref{eq:m5}. Two-stage optimization approaches, where the first objective is optimized, taking into account the hypothesis class, before the hypothesis itself is optimized for predictive accuracy (or some other such measure), have been shown to work well \cite{Johansson2016}. Secondly, such a method is specifically geared toward producing
a hypothesis that makes use of features that have an immediate bearing upon the problem, while eliminating interpretability obscuring effects, such as multi-collinearity. Moreover, these desirables are obtained while attempting to produce the most accurate hypothesis: an $\bh$ that elicits feature indicativeness, produces accurate results, and controls for confounding effects is the goal of this two-step optimization procedure.

Ultimately, we conduct our analysis by observing the sign and magnitude
of the values in the hypothesis vector in order to determine the factors
that influence hand hygiene compliance, and whether such factors
affect compliance in a positive or negative manner. We also observe
correlation and RMSE values to determine how well our
predictive model works, and whether the corresponding results can be trusted. All results and are obtained via $k$-fold cross-validation
($k=10$).

\subsubsection{Supporting Methodology}

We also use two established/standard techniques -- RReliefF feature ranking and marginal effects modeling -- that will serve as a point of comparison between our method, and also help inform the discussion of the obtained results\footnote{Note that both the LASSO \cite{Tib1996} and Elastic Net \cite{Zou2005} would have also made appropriate supporting methods.}.

\textbf{Feature ranking: }First, we propose the use of the RReliefF algorithm \cite{Robnik1997},
a modification of the original Relief algorithm of Kira and Rendell \cite{Kira1992}.
RReliefF finds a feature $j$'s weight by randomly selecting a seed instance
$\bx_i$ from design matrix $\mathbf{X}$ and then using that instance's $k$ nearest
neighbors to update the attribute. This description consists of three
terms: the probability of observing a different rate of hand hygiene
compliance than that of the current value given that of the nearest
neighbors, given by
\begin{align} 
A = p(\text{rate} \neq \text{rate}_{x_{i,j}}|k\text{NN}(x_{i,j})), 
\end{align}
the probability of observing the current attribute
value given the nearest neighbors, given by
\begin{align}
B = p(x_{i,j}|k\text{NN}(x_{i,j})),
\end{align}
and the probability of observing
a different hand hygiene rate than the current value given a different
feature value $v$ and the nearest neighbors, given by
\begin{align}
C =  p(\text{rate} \neq \text{rate}_{x_{i,j}}|k\text{NN}(x_{i,j}) \land j=v). 
\end{align}
Attribute distance
weighting is used in order to place greater emphasis on instances
that are closer to the seed instance when updating each term; final
weights are obtained by applying Bayes' rule to the three terms maintained
for each attribute, which can be expressed
\begin{align}
\frac{C*B}{A}-\frac{(1-C)*B}{1-A}. 
\end{align}
By using
this method we could then rank attributes in terms of their importance.
We again report rankings using $k$-fold ($k=10$) cross validation.

\textbf{Marginal Effects Modeling: }To provide additional insight into the features that are relevant
to hand hygiene we analyzed their marginal effects \cite{Williams2012}.
Marginal effects, also referred to as \textit{instantaneous rates of change},
are computed by first training a hypothesis $h$, then,
using the testing data, the effect of each covariate can be estimated
by holding all others constant and observing the predictions. Such a method
can be expressed by
\begin{align}
\label{eq:marg}
\hat{\text{rate}}_{i,j} = \bh^{\top}[x_{i,j},\bar{\bx}_{\neq j}]
\end{align}
where, with a slight abuse of notation, $x_{i,j}$,  the value of instance $i$'s $j$th feature, is added to the vector $\bar{\bx}_{\neq j}$, which consists of the average of each non-$j$ feature, at the appropriate location (namely, the $j$th position). Here, the notation $\neq j$ is used to reinforce the fact that the vector of averages $\bar{\bx}$ has it's $j$th element replaced by $x_{i,j}$. Other non-$j$ entries are given by $\bar{\bx}_k = \mu(\bX_k)$, for an arbitrary index position $k$.

\section{Results}

\subsection{Predictive Power: $M5$ Ridge Regression}

We learned a hypothesis using all available features, including a nominalized
facility identifier. Our predictive results can be observed in Table
\ref{tab:all_results}. We note that the RMSE is not large and the correlation is moderate, implying relatively good predictive performance.

\begin{table}[h]
	\centering
	\begin{tabular}{|l|c|}
		\hline
		\textbf{Measure} & \textbf{Value} \\
		\hline
		Correlation & 0.3441 \\
		\hline
		RMSE & 0.1702 \\
		\hline
	\end{tabular}
	
	\caption{Correlation coefficient and RMSE of cross-validated model predictions.\label{tab:all_results}}
\end{table}

\begin{table}[b]
	\centering
	\begin{tabular}{|l|c|}
		\hline
		\textbf{Feature} & \textbf{$h_j$} \\
		\hline
		\textcolor{red}{$\text{Facility}^{-}$} $= \{1, 105, 147, 156, 157, 170\}$ & $h_{j \in \text{Fac}^{-}} \in$\\
		& $[-0.103,-0.016]$ \\
		\hline
		\textcolor{blue}{$\text{Facility}^{+}$} $= \{91, 119, 123, 127, 135, 144,$ & $h_{j \in \text{Fac}^{+}} \in$ \\
		$145, 149, 153, 155, 168, 173\}$  & $[0.008, 0.261]$ \\
		\hline
		\textcolor{blue}{Air Temp} & $0.022$ \\
		\hline
		\textcolor{blue}{Rel.~Humid} & $0.0079$ \\
		\hline
		\textcolor{blue}{$weekday$} & $0.0069$ \\
		\hline
		\textcolor{red}{$nightShift$} & $-0.0218$ \\
		\hline
		\textcolor{red}{$holiday$} = $\{\text{Indep Day}, \text{Pres. Day},$ & $h_{j \in \text{Hol}}$ \\ 
		$\text{Vet Day}, \text{New Year's}, \text{Christmas}\}$ & $[-0.017,-.006]$\\
		\hline
		\textcolor{blue}{Flu Severity} & $0.014$ \\
		\hline	
		\textcolor{red}{$JulyEffect$} & $-0.0106$ \\
		\hline
	\end{tabular}
	\caption{Feature specific $h_j$ terms, where \textcolor{red}{red} highlights features with a negative association and \textcolor{blue}{blue} highlights those with a positive association. \label{tab:dn_all_res_coeff}}
\end{table}

\subsection{Examining Hypothesis $\bh^*$}

We next examine the terms of the learned hypothesis $\bh^*$ (see Table
\ref{tab:dn_all_res_coeff}). The
model includes all 19 facilities, 12 of which had positive values,
indicating relatively higher rates of compliance. The size remaining facility's $\bh^*$ terms had relatively small negative values,
indicating lower rates of compliance.
Among other features, holidays are associated with lower compliance
rates, while influenza severity has higher compliance. Weekdays
are associated with higher compliance rates, as are higher temperatures
and humidity. Interestingly, the $M5$ Ridge Regression model appears to have eliminated some holidays (Martin Luther King
day, Memorial day, Labor day, Columbus
day, and Thanksgiving), as well as Facility 163 (the facility with the lowest amount of hand-hygiene data). This means that these features do not contribute to 
hand-hygiene compliance rates in any meaningful way.

\subsection{RReliefF}

By using
RReliefF we could rank features in terms of their importance in 
order to support and supplement the result obtained using $M5$ Ridge Regression. The results are reported in Table \ref{tab:RReliefattribute-weights}, where rankings shown
are averages for 10-fold cross-validation. Note that here $facility$ was represented as a single discretely-valued feature in order to determine the importance of facility as a whole (instead of treating each facility as its own feature), as was $holiday$.

\begin{table}[h]
	\centering
	\begin{tabular}{|l|c|c|}
		\hline
		\textbf{Attribute} & \textbf{Avg Val} & \textbf{Avg Rank} \\
		\hline
		Facility & $0.029(\pm .001)$ & $1$ \\
		\hline
		Flu Sev & $0.007$ & $2$ \\
		\hline
		Air Temp & $0.005$ & $3.3(\pm 0.46)$ \\
		\hline
		$weekday$ & $0.002$ & $5$ \\
		\hline
		Rel.~Humid. & $.001$ & $6.3(\pm 0.64)$ \\
		\hline
		$JulyEffect$ & $\approx 0.0$ & $7.2(\pm 0.4)$ \\
		\hline
		$holiday$ & $\approx 0.0$ & $7.8(\pm 1.08)$ \\
		\hline
		$nightShift$ & $\approx 0.0$ & $8.7(\pm 0.46)$ \\
		\hline
		
 	\end{tabular}
	
	\caption{RReliefF attribute weights. \label{tab:RReliefattribute-weights}}
\end{table}

\subsection{Marginal Effects}

\begin{figure*}
	\centering 
	\begin{subfigure}[h]{.49\linewidth}
		\centering
		\includegraphics[width=.80\linewidth]{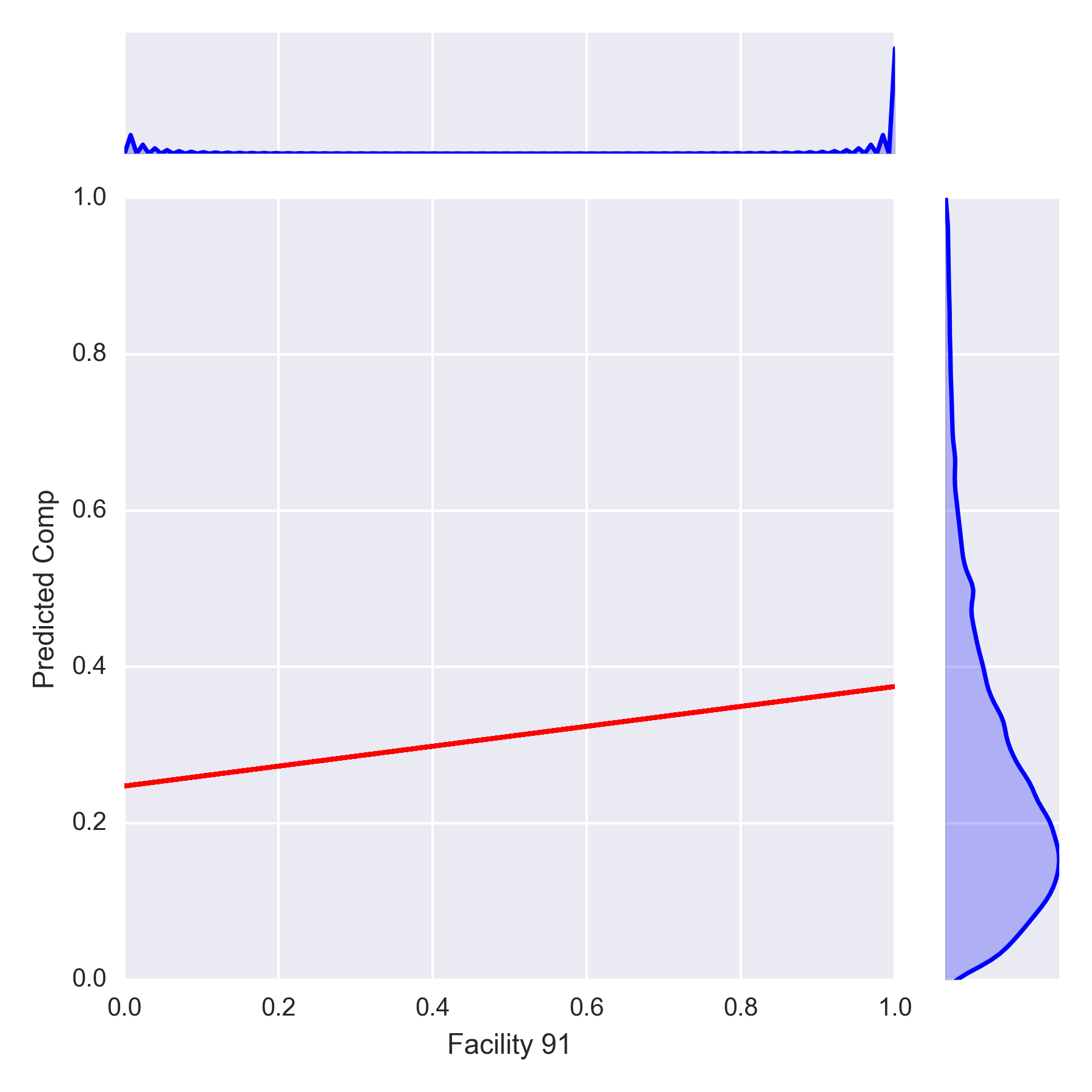}
		\caption{Facility 91. \label{fig:me91}}
	\end{subfigure}
	\begin{subfigure}[h]{.49\linewidth}
		\centering
		\includegraphics[width=.80\linewidth]{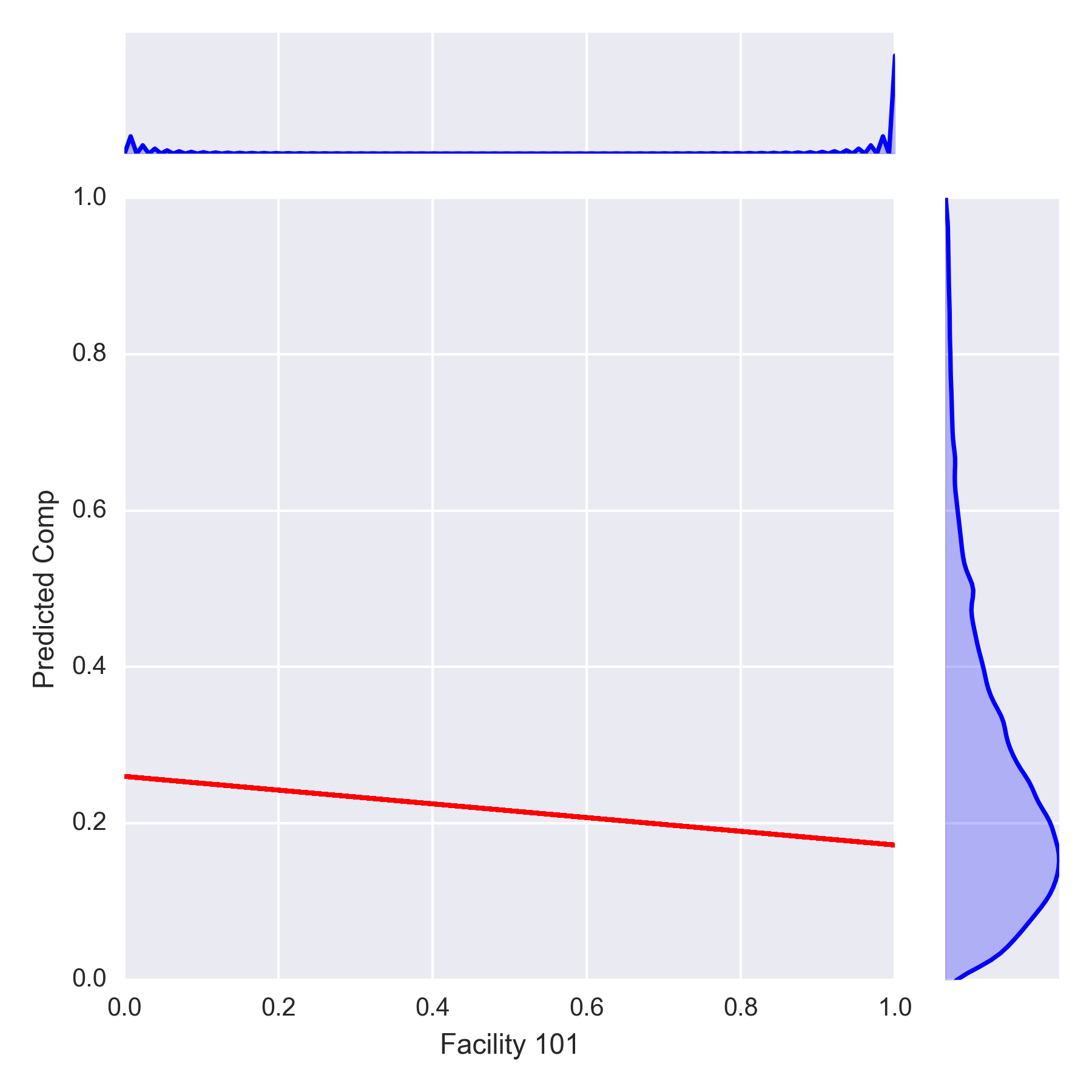}
		\caption{Facility 101.\label{fig:me101}}
	\end{subfigure}
	\begin{subfigure}[h]{.49\linewidth}
		\centering
		\includegraphics[width=.80\linewidth]{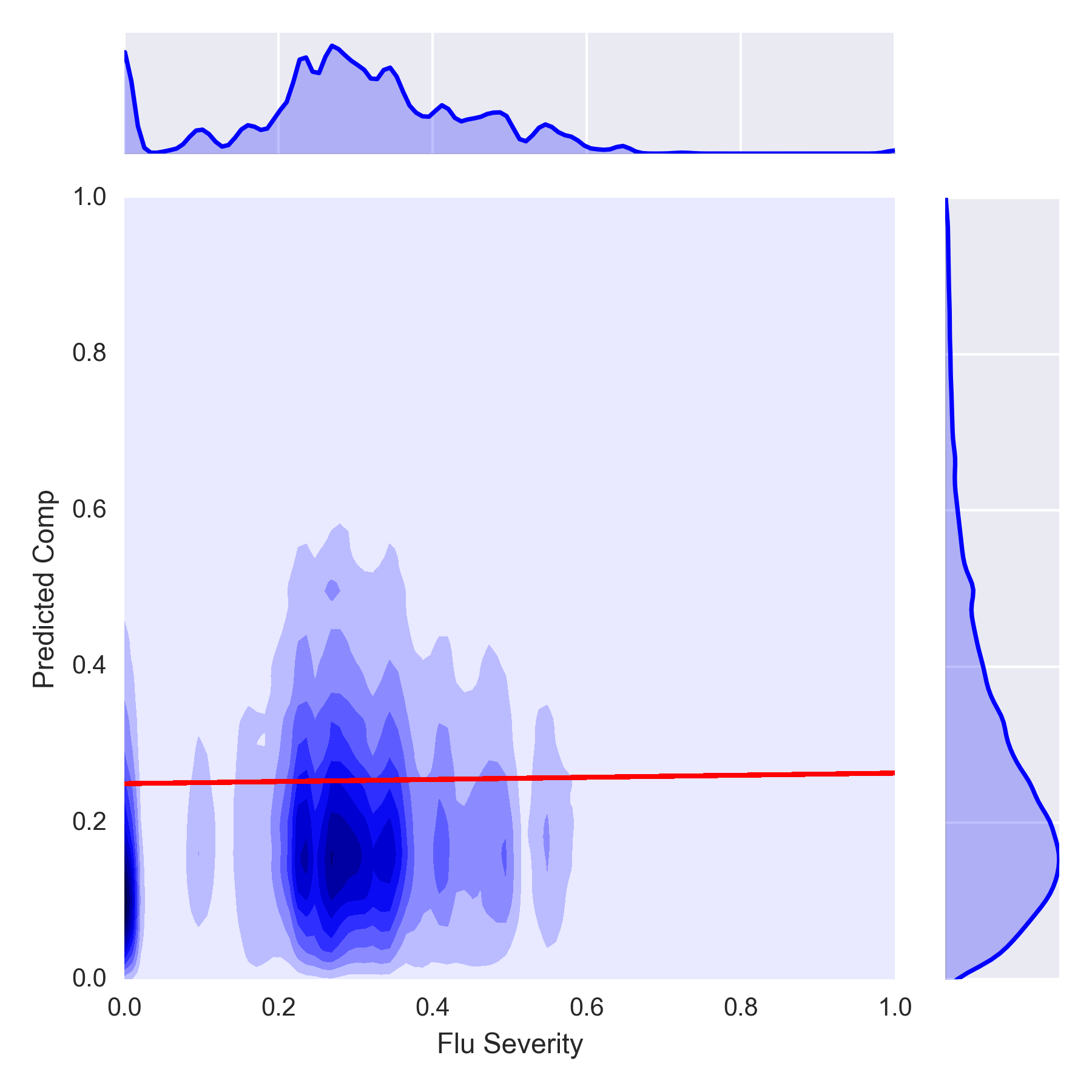}
		\caption{Flu Severity. \label{fig:mefs}}
	\end{subfigure}\par
	\begin{subfigure}[h]{.49\linewidth}
		\centering
		\includegraphics[width=.80\linewidth]{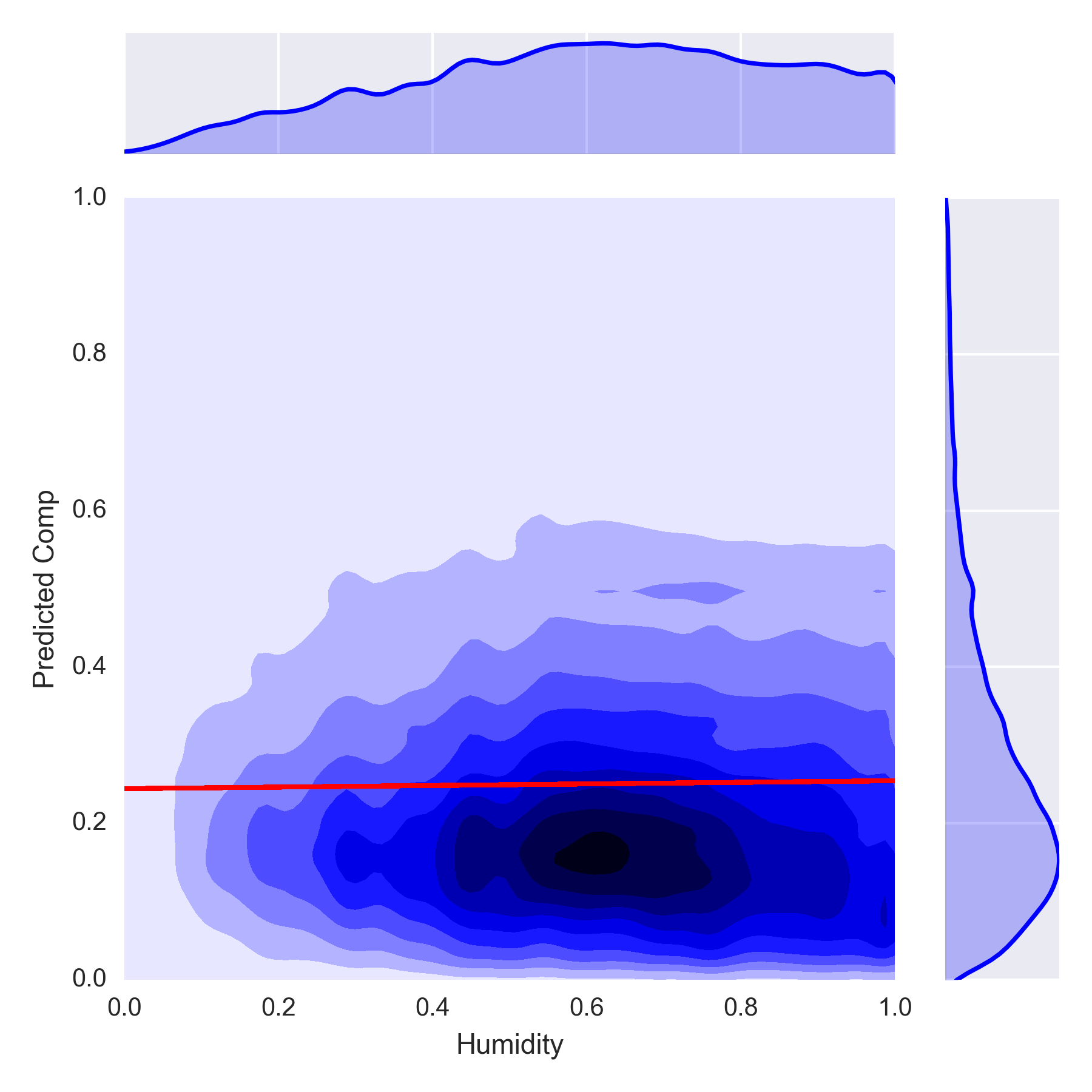}
		\caption{Humidity.\label{fig:mehum}}
	\end{subfigure}
	\begin{subfigure}[h]{.49\linewidth}
		\centering
	\includegraphics[width=.80\linewidth]{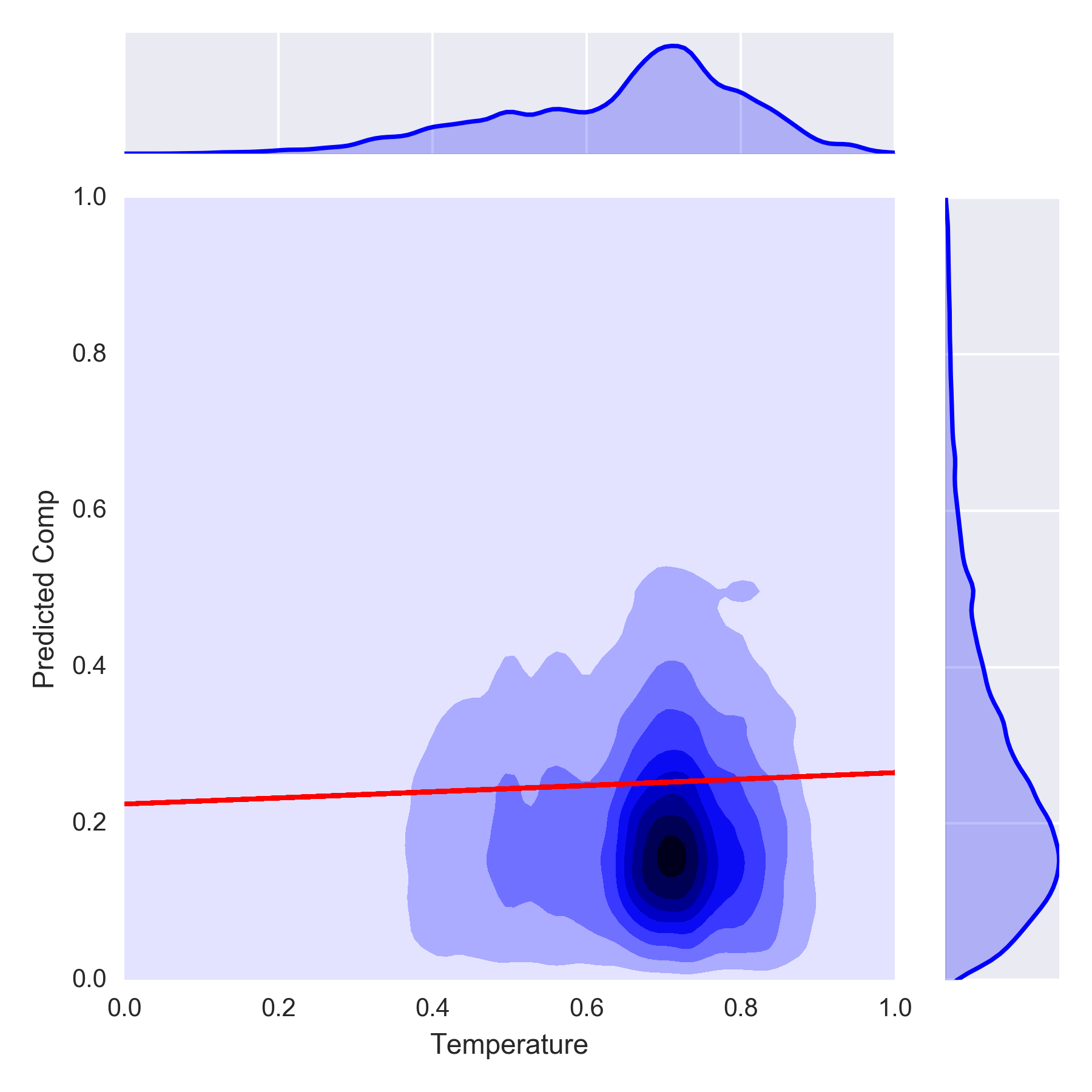}
		\caption{Temperature. \label{fig:metemp}}
	\end{subfigure}
	\caption{The marginal effects of several select covariates, where \textcolor{blue}{blue} shows the kernel density of the original data and the \textcolor{red}{red lines} show the estimation. Rate (y-axis) vs. feature (x-axis). Note that in \ref{fig:me91} and \ref{fig:me101} no kernel density estimate is provided, as these plots are for binary features.\label{fig:margef}}
\end{figure*}

The results obtained from modeling the marginal effects can be observed in Figure \ref{fig:margef}.

Figures \ref{fig:me91} and \ref{fig:me101} show the marginal effects of two randomly selected facilities; one identified as being associated with lower rates of compliance and one identified as having higher rates of compliance (from Table \ref{tab:dn_all_res_coeff}). Note that, because these are binary features (taking on values of either zero or one), the kernel density of the underlying data is not readily visible (unlike the other figures, which show results for non-binary features). As we can see the marginal effects support the result obtained using both $M5$ Ridge Regression and RReliefF, and also seem to suggest an even greater association between facilities and rates of compliance than was originally apparent (at least for these two facilities).

Figure \ref{fig:mefs} shows the marginal effects of flu Severity. The Flu Severity result shows a slightly positive relationship between the severity of flu, measured in terms of mortality, and hand-hygiene compliance rates. This is further supported by the result obtained from $M5$ Ridge Regression and the RReliefF ranking.

Figures \ref{fig:mehum} and \ref{fig:metemp} show the marginal effects of humidity and temperature. The result obtained for both is consistent with that from $M5$ Ridge Regression. The lesser effect of humidity and greater effect of temperature are also reflected in the RReliefF ranking.

To further explore the relationship between hand-hygiene and weather effects, we conducted a simple statistical analysis. For each facility, we selected the temperature and humidity values corresponding to the bottom 10\% and top 10\% of hand-hygiene compliance rates. We then performed a paired t-test on each set of samples; temperature and humidity values were scaled to $[0,1]$. The results of this analysis are reported in Table \ref{tab:ttest}.

\begin{table}[h]
	\centering
	\begin{tabular}{|l|c|c|c|}
		\hline
		\textbf{Facility} & \textbf{State} & \textbf{Temperature} & \textbf{Humidity} \\
		& & $\mu_{\text{top}} - \mu_{\text{bot}}$ (p-val) & $\mu_{\text{top}} - \mu_{\text{bot}}$ (p-val)\\
		\hline
		91 & OH & -0.004 (0.750) & -0.007 (0.489)  \\
		\hline
		101 & OH & 0.001 (0.909) & 0.004 (0.457)  \\
		\hline
		\textcolor{blue}{105} & TX & 0.041 ($< 0.000$) & -0.028 (0.001) \\
		\hline
		119 & MN & -0.008 (0.699) & -0.013 (0.337) \\
		\hline
		\textcolor{blue}{123} & TX & 0.017 (0.002) & 0.029 ($< 0.000$) \\
		\hline
		\textcolor{blue}{127} & NM & 0.032 ($< 0.000$) & -0.063 ($< 0.000$) \\
		\hline
		135 & OH & -0.045 (0.010) & 0.017 (0.278) \\
		\hline
		\textcolor{blue}{144} & CA & 0.009 ($< 0.000$) & -0.018 (0.002) \\
		\hline
		145 & CA & -0.001 (0.675) & 0.004 (0.549) \\
		\hline
		\textcolor{blue}{147} & CA & 0.011 ($< 0.000$) & -0.013 (0.017)  \\
		\hline
		149 & CA & -0.007 (0.025) & 0.008 (0.214)  \\
		\hline
		\textcolor{blue}{153} & CT & 0.043 ($< 0.000$) & -0.003 (0.746) \\
		\hline
		\textcolor{blue}{155} & NY & 0.093 ($< 0.000$) & 0.012 (0.341)  \\
		\hline
		\textcolor{blue}{156} & NC & 0.040 (0.007) & -0.041 (0.445)  \\
		\hline
		157 & OH & -0.132 ($< 0.000$) & -0.020 (0.638)  \\
		\hline
		\textcolor{blue}{163} & OH & 0.180 (0.010) & 0.179 (0.021) \\
		\hline
		\textcolor{blue}{168} & PA & 0.012 (0.122) & 0.071 (0.006)  \\
		\hline
		170 & IL & -0.001 (0.772) & -0.007 (0.642) \\
		\hline
		\textcolor{blue}{173} & OH & 0.037 (0.003) & -0.033 (0.440) \\
		\hline
	\end{tabular}
	\caption{The difference in means and paired t-test p-value results, obtained by comparing temperature/humidity values among the bottom 10\% and top 10\% of hand-hygiene compliance rates, by facility (\textcolor{blue}{blue} indicates that either temperature, humidity, or both have a positive difference in means and a p-value $\leq .05$).\label{tab:ttest}}
\end{table}

Table \ref{tab:ttest} shows that most facilities have statistically significant differences between the two samples and that $\mu_{\text{top 10}} > \mu_{\text{bottom 10}}$. Such results indicates that higher temperatures and levels of humidity (particularly temperature) are statistically associated with higher rates of hand hygiene. However, we find that some facilities co-located in the same geographic region have conflicting statistical results (e.g.,~ Facs.~91, 173). We conjecture that such a result may attributable to differences in sensor deployment location, but we leave such an investigation as future work.

\subsection{Facility-Specific Modeling}

The full $M5$ Ridge Regression models' reliance on facility identities suggests that
compliance relies, at least in part, on facility-specific health care
worker attitudes, administrative culture, or even simply the disposition
of sensors and the architecture of the facility. Given the magnitude
of the coefficients associated with facilities in
the previous model, we propose to construct and analyze a facility-specific model.

Here, we selected a facility (facility 91) with both a high rate of
compliance and a large number of reported events for further investigation
(see Table \ref{tab:facility_91_res}). As expected, the facility-specific model is better at predicting compliance than the full model (Table \ref{tab:all_results}), while the correlation is comparable.

\begin{table}[h]
	\centering
	\begin{tabular}{|l|c|}
		\hline
		\textbf{Measure} & \textbf{Value} \\
		\hline
		Correlation & 0.3179 \\
		\hline
		RMSE & 0.0381 \\
		\hline
	\end{tabular}

	\caption{Correlation coefficient and RMSE of a cross-validated model for Facility 91. \label{tab:facility_91_res}}
\end{table}

The hypothesis terms associated with this model are shown in Table \ref{tab:facility_91_coeff}.
Unlike the previous model, temperature is negatively associated
with compliance, which is somewhat surprising. We also note a larger negative association between compliance and
flu severity which, while somewhat harder to explain, may also reflect
the narrower geographic scope accounted for by this model. Ultimately, only $weekday$ and humidity positively impact compliance, which is a different result than in our global model. These differences aren't surprising, however: the original
model attempts to capture effects across a broad geographic region,
while this model need only capture the associations found in a specific location.

\begin{table}[h]
	\centering
	\begin{tabular}{|l|c|}
		\hline
		\textbf{Feature} & \textbf{$h_j$} \\
		\hline
		\textcolor{red}{Air Temp} & $-0.0858$ \\
		\hline
		\textcolor{blue}{Rel.~Humid.} & $0.0546$ \\
		\hline
		\textcolor{blue}{Weekday} & $0.039$ \\
		\hline
		\textcolor{red}{Day Shift} & $-0.1742$ \\
		\hline
		\textcolor{red}{Flu Sev.} & $-0.2097$ \\
		\hline
	\end{tabular}

	\caption{Feature specific $h_j$ terms for the Facility 91 model, where \textcolor{red}{red} highlights features with a negative association and \textcolor{blue}{blue} highlights those with a positive association. \label{tab:facility_91_coeff}}
\end{table}

\section{Discussion and Future Work}

In this section we discuss the broader implications of our findings, as well as directions for future work. 

%
%
The full model and marginal effects models, in conjunction with the RReliefF feature ranking, provided several insights. First, we found that facility identities are strongly related
to compliance, suggesting that facility-wide attitudes
towards hand hygiene exist, persist in time, and are predictive of
compliance rates. This observation may also reflect differences in sensor installation, where different facilities may have sensors instrumented in different departments, thus affecting reported rates. 
Second, increases in influenza severity were associated with an increase in compliance, which is encouraging.
Third, our conjecture regarding lower weekend and holiday compliance appears to have some merit, although the holidays associated with negative compliance were somewhat surprising. We again acknowledge that this result may be affected by increased visitors during these times. 
Fourth, our conjectures that higher humidity and temperature are indicative
of higher rates of compliance were confirmed by the full model, marginal effects model, and statistical analysis. This finding is important as health care workers often cite skin irritation or dry skin as reasons for reduced frequency of hand hygiene. 
Fifth, we found that compliance during the first week
of residents' attendance ran contrary to our original conjecture: the $JulyEffect$ was essentially unobservable.
Finally, we found that $nightShift$ was associated with slightly lower compliance rates.

Our facility-specific model (constructed for Facility 91) found contradictions with the full hypothesis. We believe that this supports the facility-specific attitudes conjecture and that, moreover, different factors may be at play at different facilities spanning different geographical regions. Further work is needed to tease these differences out, however.

This work has several limitations. First, there are differences among
installations: not all doors and dispensers may be instrumented and, therefore, we cannot track, for example, the use of personal alcohol
dispensers (we assume stable practices). Thus our compliance estimates may be based on partial
information and are certainly not comparable across facilities. Second, our compliance estimates are facility wide, meaning
that we do not exploit the co-location of dispensers and door event
sensors, but only the temporal correlation of the individual events.
Thus, our assumption that each door event corresponds to a hand-hygiene
opportunity may be fundamentally flawed, even as it allows for consistent
intra-facility comparisons. Third, we acknowledge the possibility
of location and sampling bias with regard to both the sensors and
facilities. If sensors were to be placed in only the ICU of one facility
and in the emergency room of another, we may observe different
rates, which has not been accounted for. Additionally, though facilities
are distributed across the United States, they are by no means meant
to be a representative sample of facility types or climatic conditions.

There are also a number of opportunities for future work. First, we would like to consider alternative definitions of compliance and examine compliance at finer-grained temporal levels, perhaps incorporating time-series analyses as an additional avenue of exploration. We intend to 
also explore framing the problem as one of classification, rather than only regression, which may help tease out uncertain factors. Finally, data pertaining to compliance rates under certain interventions
would give way to exploration of intervention efficacy both in general and using prediction-based methodology, such as inverse classification, to recommend facility-specific intervention policies \cite{Lash2016,Lash2016b}.


Hand hygiene compliance is a simple yet effective method of preventing
the transmission of disease, both among the population at large, and
within health care facilities. This study presents a first look at
factors that underlie health care worker hand-hygiene compliance rates,
including weather conditions, holidays and weekends, and infectious
disease prevalence and severity, and serves as a model for future
studies that will exploit the availability of temporally and spatially
rich compliance data collected by the sophisticated sensor systems
now being put into practice.

\section*{Acknowledgments}

The authors would like to thank Gojo Industries for access to the
hand-hygiene data and for their financial support of this work, as
well as Andrew Arthur for his help in preparing the data.



%
\bibliographystyle{IEEEtran}
\bibliography{ichi_15_hand}

\end{document}